\documentclass[graybox]{svmult}

\usepackage{mathptmx}       
\usepackage{helvet}         
\usepackage{courier}        
\usepackage{type1cm}        
\usepackage{makeidx}         
\usepackage{graphicx}        
\usepackage{multicol}        
\usepackage[bottom]{footmisc}
\usepackage{epsfig}
\usepackage{subfig}
\graphicspath{
               {/home/remy/Dropbox/book_chapter/4publication/}
              }

\makeindex             

\begin{document}

\title*{Voltage interval mappings for an elliptic bursting model}
\author{Jeremy Wojcik and Andrey Shilnikov}
\institute{Jeremy Wojcik \at Neuroscience Institute, and
Department of Mathematics and Statistics,
Georgia State University, Atlanta, 100 Piedmont Ave SE, Atlanta, GA, 30303, USA \email{jwojcik1@gsu.edu}
\and Andrey Shilnikov \at Neuroscience Institute, and
Department of Mathematics and Statistics,
Georgia State University, Atlanta,  100 Piedmont Ave SE, Atlanta, GA, 30303, USA \email{ashilnikov@gsu.edu}}
%
%
\maketitle


\abstract{We employed Poincar\'e return mappings for a parameter interval to an exemplary elliptic bursting model, the FitzHugh-Nagumo-Rinzel model.
 Using the interval mappings, we were able to examine in detail  the bifurcations that underlie the complex
activity transitions between: tonic spiking and bursting, bursting and mixed-mode oscillations, and finally,
mixed-mode oscillations and quiescence
in the FitzHugh-Nagumo-Rinzel model. We illustrate the wealth of information, qualitative and quantitative, that was derived from the Poincar\'e
mappings, for the neuronal models and for similar (electro)chemical systems.}

\section{Introduction}
\label{sec:1}

The class of elliptic bursting models is rich and can be found in diverse scientific studies, ranging from biological systems \cite{wojcik_1d} to chemical processes
such as the Belousov-Zhabotinky reaction \cite{argoul}. Transitions between activity states for elliptic bursting models is not common knowledge. Often in the sciences
specialization  leads to discoveries  that remain unknown in other branches of science; the recent reincarnation of mixed mode oscillations (MMO)
in neuroscience for example. In neuroscience, transitions in activity revolve around a changing membrane potential and specific changes in potential may instigate the onset
of a seizure in the case of epilepsy or determine muscle reactions in response to stimulus. The class of elliptic bursting models needs a
more general treatment that can span multiple disciplines.  We propose a case study of the phenomenological FitzHugh-Nagumo-Rinzel model in order to investigate
the mechanisms for state transitions in dynamic behavior.

Bursting represents direct evidence of multiple time scale dynamics of a model.
Deterministic modeling of bursting models was originally proposed and done within
a framework of three-dimensional, slow-fast dynamical systems. Geometric configurations
of models of bursting neurons were pioneered by Rinzel~\cite{Rinzel87b,Rinzel1987} and
enhanced in \cite{Bertram1995,Guckenheimer1996}. The
proposed configurations are all based on the geometrically comprehensive dissection approach
or the time scale separation which has become the primary tools in mathematical neuroscience.
The topology of the slow motion manifolds is essential to the geometric understanding of dynamics.
 Through the use of geometric methods of
the slow-fast dissection, where the slowest variable of the model is treated as a control
parameter, it is possible to detect and follow the manifolds made of branches of equilibria and
limit cycles in the fast subsystem. Dynamics of a slow-fast system are determined by,
and centered around, the attracting sections of the slow motion manifolds \cite{Arnold1994,Mi1994,Ti,Neishtadt1988}.

The slow-fast dissection approach works exceptionally well for a multiple time scale model,
provided the model is far from a bifurcation in the singular limit. On the other hand, a
bifurcation describing a transition between activities may occur
from reciprocal interactions involving the slow and fast dynamics of the model. Such slow-fast interactions may lead to the emergence of
distinct dynamical phenomena and bifurcations that can  occur only in the full
model, but not in either subsystem of the model. As such, the slow-fast dissection fails
at the transition where the solution is no longer constrained to stay near the slow motion manifold,
or when the time scale of the dynamics  of the fast subsystem slows to that of the slow system, near
the homoclinic and saddle node bifurcations for example.

Transformative
bifurcations of repetitive oscillations, such as bursting, are most adequately described by Poincar\'e
mappings \cite{Shilnikov2001}, which allow for global bifurcation analysis. Time series
based Poincar\'e mappings have been heavily employed for examinations of voltage oscillatory
activities in a multitude of applied sciences
\cite{Albahadily1989,Gaspard1987,Hudson1979}, despite their limitation
due to sparseness. Often, feasible
reductions to mappings of the slowest variable can be achieved through the
aforementioned dissection tool in the singular limit \cite{Griffiths2006,Medvedev2006,Shilnikov2008a,Shilnikov2001}.
However, this method often fails for elliptic bursters since no single valued mapping for the slow variable can be derived
for the particular slow motion manifold.

In this paper, we refine and expound on the technique of creating a family of one-dimensional mappings,
proposed in \cite{Channell2007a,Channell2007,Channell2009}, for the leech heart interneuron,
into the class of elliptic models of endogenously bursting neurons.  We will show that a plethora of information,
both qualitative and quantitative can be derived from the mappings to thoroughly describe
 the bifurcations as
such a  model undergoes transformations.  We also demonstrate the power
of deriving not only individual mappings, but the additional benefits of
having the entire family of mappings created from an elliptic bursting
model.

\section{FitzHugh-Nagumo-Rinzel Model}
\label{sec:2}

\begin{figure}[b!]
\begin{center}
\centerline{\includegraphics[width=0.5\textwidth]{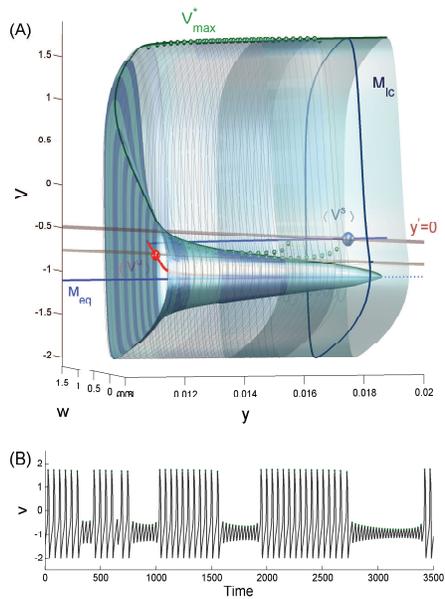}}
\end{center}
\caption{(A) Topology of the tonic spiking, $\mathrm{M_{lc}}$, and quiescent,
$\mathrm{M_{eq}}$ manifolds. Solid and dashed branches of $\mathrm{M_{eq}}$ are made of stable and unstable equilibria of the model, resp. The space curve,
labeled by $\mathrm{V^{*}_{max}}$ (in green), corresponds to the v-maximal coordinates of the periodic orbits
composing $\mathrm{M_{lc}}$.  An intersection point of $y^\prime=0$ with $\mathrm{M_{eq}}$ is an equilibrium state of (\ref{fhr}).
Shown in grey is the bursting trajectory traced by the phase point: the number of spikes per burst is the same as the number
of turns the phase point makes
around ${\mathrm M_{lc}}$.  Spikes are interrupted by the periods of quiescence when the phase point follows
$\mathrm{M_{eq}}$ after it falls from $\mathrm{M_{lc}}$ near the fold. (B) A voltage trace for $c=-0.67$ displaying
the voltage evolution in time as the phase point travels around the slow motion manifolds. }\label{fig1}
\end{figure}

We introduce the exemplary phenomenological elliptic bursting model, the FitzHugh-Nagumo-Rinzel model. The model exhibits all necessary traits for the class of elliptic bursters:
the time series form elliptic shaped bursts and oscillations are begin through an Andronov-Hopf bifurcation and end in a saddle-node bifurcation.
The model exhibits several types of oscillations including: constant high amplitude oscillatory behavior (tonic spiking), bursting, low amplitude oscillations, and MMO.
The mathematical FitzHugh-Nagumo-Rinzel model of the elliptic burster is given by the following system of equations with a single cubic nonlinear term:
 \begin{equation} \label{fhr}
 \begin{array}{rclcl}
 v^\prime &= & v-v^3/3-w+y+I,\\
 w^\prime &= & \delta(0.7+v-0.8w),\\
 y^\prime &= & \mu(c-y-v);
 \end{array}
\end{equation}
here we fix $\delta=0.08$, $I=0.3125$ is an applied external current, and  $\mu=0.002$ is a small
parameter determining the pace of the slow $y$-variable. The slow variable, $y$, becomes frozen in the singular limit, $\mu=0$. We employ c as the
primary bifurcation parameter of the model, variations of which elevate/lower the slow nullcline given by $y^\prime=0$.
The last equation is held geometrically in a plane given by $v=y-c$ in the three-dimensional
phase space of the model, see Fig.\ref{fig1}. The two fast equations in (\ref{fhr}) describe a relaxation
oscillator in a plane, provided $\delta$ is small.
The fast subsystem exhibits either tonic spiking oscillations or quiescence for different values
of y corresponding to a stable limit cycle and a stable equilibrium state, respectively.
The periodic oscillations in the fast subsystem are caused by a
hysteresis induced by the cubic nonlinearity in the first ``voltage" equation of the model.

Fig.~\ref{fig1}~(A) presents a 3D view of the slow motion manifolds in the phase space of the FitzHugh-Nagumo-Rinzel model.
The tonic spiking manifold $\mathrm{M_{lc}}$ is composed of the limit cycles for the model (\ref{fhr}),
both stable (outer) and unstable (inner) sections. The fold on $\mathrm{M_{lc}}$ corresponds to a saddle-node bifurcation, where
the stable and unstable branches merge. The vertex, where the unstable branch of
$\mathrm{M_{lc}}$  collapses at $\mathrm{M_{eq}}$,  corresponds to a subcritical Andronov-Hopf bifurcation.
The manifold $\mathrm{M_{eq}}$ is the space curve made from equilibria of the model. The intersection of the plane, $y^\prime=0$
with the manifold,
determines the location of the existing equilibrium state for a given value of the bifurcation parameter $c$:
stable (saddle-focus) if located before (after) the Andronov-Hopf bifurcation point on the solid (dashed) segment of
$\mathrm{M_{eq}}$.  The plane, $y^\prime=0$, called the slow nullcline, above (below) which the $y$-component of a solution
of the model increases (decreases).  The plane moves in the 3D phase space as the control parameter $c$ is varied.
When  the slow nullcline cuts through the solid segment of $\mathrm{M_{eq}}$,
the model enters a quiescent phase corresponding to a stable equilibrium state.  Raising the plane to intersect the
unstable (inner) cone-shaped portion of $\mathrm{M_{lc}}$ makes the equilibrium state unstable through the Andronov-Hopf bifurcation,
which is subcritical in the singular limit, but becomes supercritical at a given value of the small parameter
$\varepsilon=0.002$, see Fig.~\ref{fig1}(A). Continuing to raise the slow nullcline by increasing $c$ gives rise to bursting
represented by solutions  following and repeatedly switching between $\mathrm{M_{eq}}$ and  $\mathrm {M_{lc}}$.
Bursting occurs in the model (\ref{fhr}) whenever the quiescent $\mathrm {M_{eq}}$ and spiking $\mathrm {M_{lc}}$ manifolds contain
no attractors, i.e. neither a stable equilibrium state nor a stable periodic orbit exist. The number of complete
revolutions, or ``windings", of the phase point around $\mathrm {M_{lc}}$ corresponds to the number of spikes per burst.
The larger the number of revolutions the longer the active phase of
the neuron lasts. Spike trains are interrupted by periods of quiescence while the phase point follows the branch
$\mathrm{M_{eq}}$, onto which the phase point falls from $\mathrm{M_{lc}}$ near the fold, see Fig. \ref{fig1}. The length of the
quiescent period, as well as the delay  of the stability loss (determined mainly, but not entirely,
by the small parameter  $\mu$) begins after the phase point passes through  the subcritical
Andronov-Hopf bifurcation onto the unstable section of  $\mathrm{M_{eq}}$.
Further increase of the bifurcation parameter, $c$, moves the slow nullcline up so that it cuts through the stable cylinder-shaped
section of the manifold $\mathrm{M_{lc}}$ far from the fold. This gives rise to  a
stable periodic orbit corresponding to tonic spiking oscillations in the model.

\section{Voltage interval mappings}
\label{sec:2}

\begin{figure}[b!]
\begin{center}
\centerline{\includegraphics[width=0.5\textwidth]{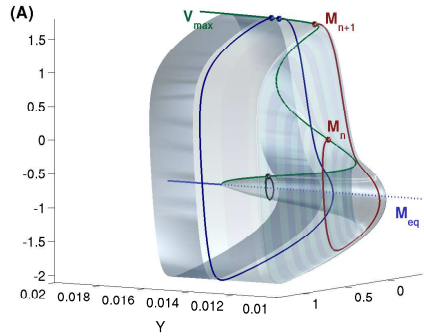}}
\end{center}
\caption{Three sample orbits demonstrating the construction of the return mapping $T$: $\mathrm{M_n} \to M_{n+1}$
defined for the points of the cross-section $\mathrm{V_{max}}$ on the manifold $\mathrm{M_{lc}}$. Singling out
the $v$-coordinates of the points gives pairs ($ \mathrm{V_{n},\, V_{n+1}}$) constituting the voltage interval mapping at a given parameter, $c$.}\label{fig2}
\end{figure}

Methods of the global bifurcation theory are organically suited for examinations of recurrent dynamics such as
tonic spiking, bursting  and subthreshold oscillations \cite{Doi2005,Gaspard92}, as well as their transformations.
The core of the method is a reduction to, and derivation of, a low dimensional Poincar\'e return mapping with
an accompanying analysis of the limit solutions: fixed, periodic and homoclinic orbits each representing various oscillations
in the original model.
and referenced therein. It is customary that such a mapping is sampled from time series, such as identification of voltage maxima, minima, or interspike intervals \cite{Gaspard1992a},
Fig.~\ref{fig1}(B).
A drawback of a mapping generated by time series is sparseness as the construction algorithm reveals only a single periodic attractor of a model, unless the latter demonstrates chaotic or mixing dynamics producing a large set of densely wandering
points. Chaos may also be evoked by small noise whenever the dynamics of the model are sensitively vulnerable to
small perturbations that do not  substantially re-shape intrinsic properties
of the autonomous model \cite{Channell2009,Su2004}. Small noise, however,  can make the solutions
of the model wander thus revealing the mapping graph.

A computer assisted method for constructing  a complete family of Poincar\'e mappings
for an interval of membrane potentials was proposed in
\cite{Channell2007} following \cite{Shilnikov1993}.
Having a family of such mappings we are able to elaborate on
various bifurcations of periodic orbits, examine bistability of coexisting tonic spiking and bursting, and detect
the separating unstable sets that are the organizing centers of complex  dynamics in any model.
 Examination of the mappings will help us
make qualitative predictions about transitions {\em before} the transitions occur in models.

By construction, the mapping $T$ takes the space curve $\mathrm{V^{*}_{max}}$ into itself after a single
revolution around the manifold $\mathrm{M_{lc}}$, Fig. \ref{fig2}, i.e.  $T:\,\mathrm{V_n} \to \mathrm{V_{n+1}}$. This technique
allows for the creation of a Poincar\'e return mapping; taking an interval of the voltage values into itself. The
found set of matching pairs $\mathrm{(V_{n},\,V_{n+1})}$ constitutes the graph of the Poincar\'e mapping for a
selected parameter value $c$. Provided the number of paired coordinates is sufficiently large and by applying a standard
spline interpolation we are able to iterate trajectories of the mapping, compute Lyapunov exponents,
evaluate the Schwarzian derivative, extract kneading invariants for the topological entropy, and determine many other quantities.

Varying the parameter, $c$, we are able to obtain a dense family that covers all behaviors,
bifurcations and transitions of the model (\ref{fhr}). A family of the  mappings for the parameter,
$c$, varied within the range $[-1, \,-0.55]$ is shown in
Fig.~\ref{fig3}. Indeed, for the sake of visibility, this figure depicts a sampling of mappings that
indicate evolutionary tendencies of the model. A thorough examination of the family allows us to foresee changes in model dynamics.
A family of mappings allows us to analyze all the
bifurcations whether stable or unstable fixed and periodic orbits including homoclinic and heteroclinic
orbits and bifurcations. By
following the mapping graph we can predict a value of the parameter at which the corresponding periodic orbit
will lose stability or vanish, for example giving rise to bursting from tonic spiking.

A fixed point, $v^{\star}$, is discerned from the mapping as an intersection of the graph
with the bisectrix. Visually we determine the stability of the fixed point by the slope of the graph at the fixed point. If
the slope of the graph is less than $1$ in absolute value the point is stable. When the absolute value of the slope of the graph at
the fixed point is greater than $1$ the fixed point is unstable.
Alternatively stability may be determined from forward iterates of an initial point in the neighborhood of the fixed point which
converges to the fixed point.

\section{Qualitative analysis of mappings}

\begin{figure}[b!]
\begin{center}
\centerline{\includegraphics[width=0.5\textwidth]{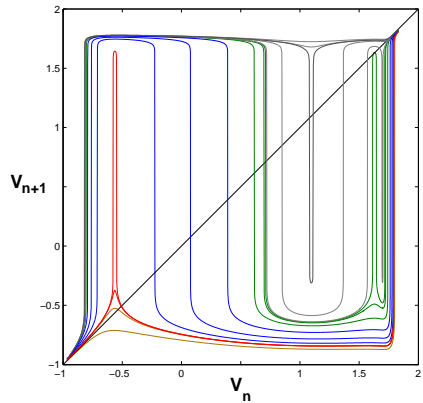}}
\end{center}
\caption{Coarse sampling of the $c$-parameter family of the Poincar\'e return mappings $T: \mathrm{V_n} \to \mathrm{V_{n+1}}$ for
the FitzHugh-Nagumo-Rinzel model at $\mu=0.002$ as $c$ decreases from  $c=-0.55$ through $c=-1$.
The grey mappings correspond to the dominating tonic spiking activity in the model. The green mappings show the model transitioning
from tonic spiking to bursting. The blue mappings correspond to the bursting behavior in the model. The red mappings
show the transition from bursting into quiescence. The orange mappings correspond to the quiescence in the model. An intersection point of a mapping graph with the
bisectrix is a fixed point, $v^{\star}$, of the mapping. The stability of the fixed point is determined by the slope of the mapping graph,
i.e. it is stable if $|T'(v^{\star})|<1$. Nearly vertical slopes of graph sections are
due to an exponentially fast rate of instability of solutions (limit cycles) of the fast subsystem  compared to
the slow component of the dynamics of the model.}\label{fig3}
\end{figure}

The family of mappings given in Fig.~\ref{fig3} allows for global evolutionary tendencies of the model~(\ref{fhr}) to be
qualitatively analyzed. One can first see that the flat mappings in grey have a single fixed point corresponding to
the tonic spiking state. The green mappings show the actual transition and saddle-node bifurcation after
which we have regular bursting patterns, seen in the blue mappings. We also see the other unstable fixed point clearly moving to the
lower corner. The red mappings indicate the transition from bursting to quiescence, as the fixed point changes stability.

A major benefit of using the voltage interval mapping is that we are able to understand transitions between the
activity states of the model by analyzing and comparing the bifurcations between the states. Activity transitions
commonly occur in a slow-fast model near the bifurcations of the fast subsystem where the description of dynamics in
the singular limit is no longer accurate because of the failure of (or interpretation of) the slow-fast dissection paradigm.
This happens, for example, when the two-dimensional fast subsystem of the model (\ref{fhr}) is close to a saddle-node
bifurcation (near the fold on the tonic spiking manifold $\mathrm{M_{lc}}$) where the fast dynamics slow to the
time scale of the slow subsystem. Such an interaction may cause new and peculiar phenomena such as torus formation and
subsequent breakdown near the fold on the spiking manifold \cite{Kramer2008a,Shilnikov2003}.
 We now turn our attention to a more thorough analysis of the individual mappings.

\subsection{Transition from tonic spiking to bursting}\label{ts2b}

\begin{figure}[b!]
\begin{center}
\centerline{\includegraphics[width=0.5\textwidth]{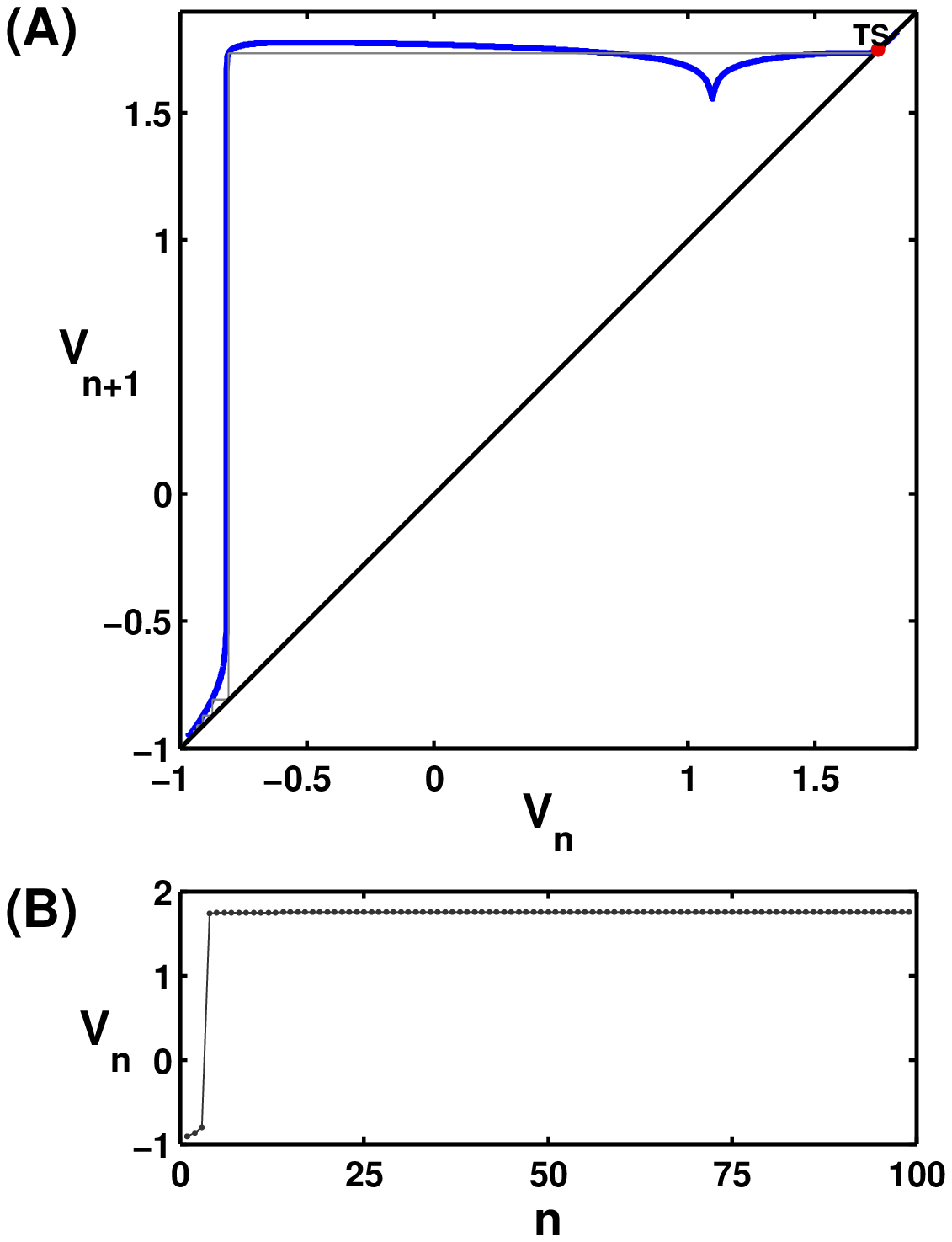}~~\includegraphics[width=0.5\textwidth]{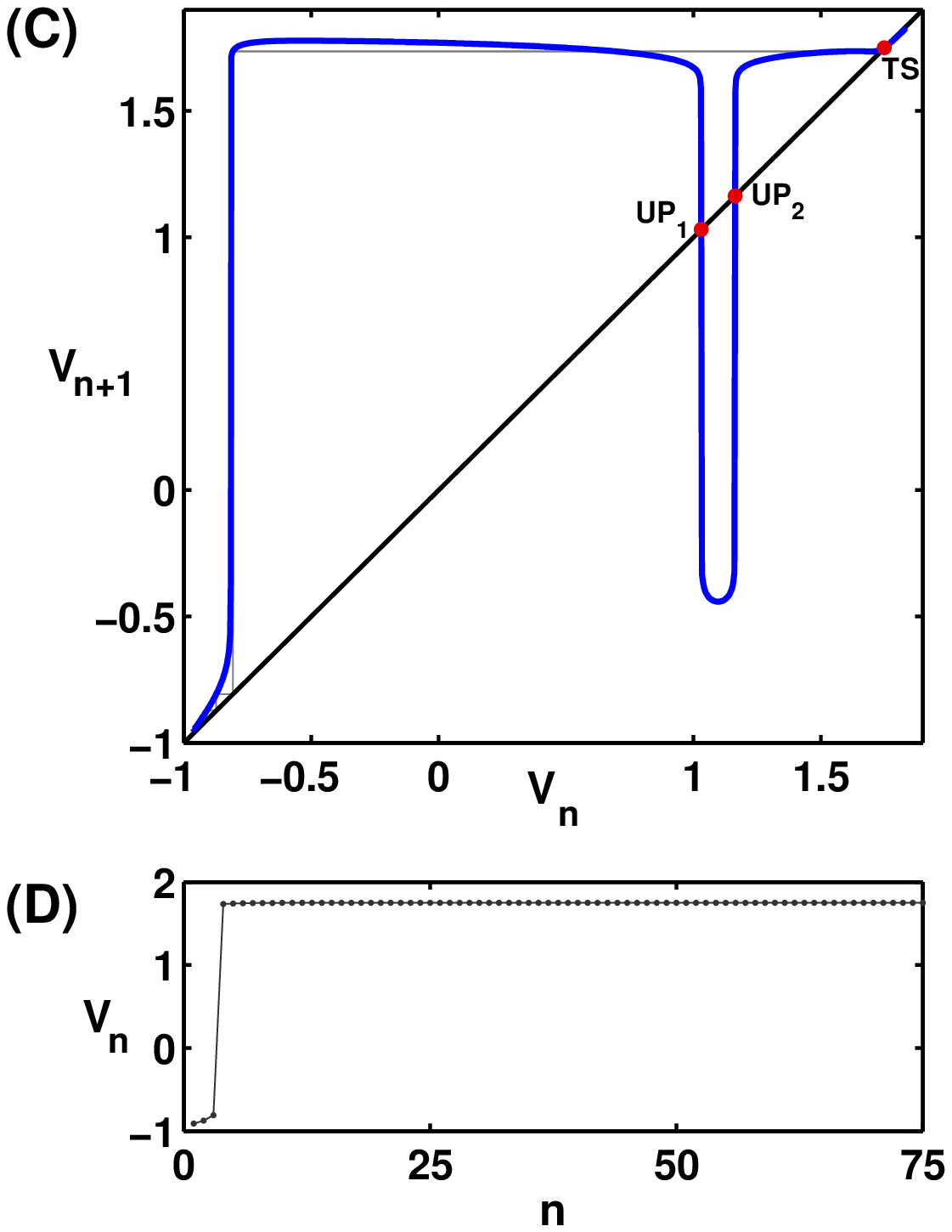}}
\end{center}
\caption{(A) Poincar\'e return mapping for the parameter, $c=-0.594255$. We see a single fixed point, TS, corresponding to continuous large amplitude
oscillations. We also see a cusp which insinuates a possible change in the mapping shape. (B) A maximal ``time'' series obtained from iterating the mapping, $n$ times.
(C) Return mapping for $c=-0.595$. We see the cusp has enlarged and intersected the identity line creating 2 additional fixed points, $UP_1$ and $UP_2$.
The two fixed points are clearly unstable. (D) There is no indication in the maximal trace, or model dynamics, that would indicate the formation
of these fixed points. }\label{fig4}
\end{figure}

\begin{figure}[b!]
\begin{center}
\centerline{\includegraphics[width=0.5\textwidth]{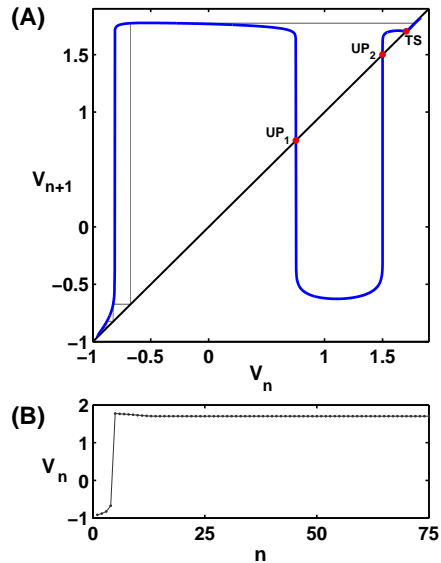}}
\end{center}
\caption{(A) Varying the parameter further to $c=-0.615$ we find the unstable fixed point $UP_2$ has moved closer to the stable fixed point, TS.
The other unstable fixed point $UP_1$ remains in approximately the same location. (B) Again the maximal trace shows no indication of any change in dynamics.}\label{fig5}
\end{figure}

\begin{figure}[b!]
\begin{center}
\centerline{\includegraphics[width=0.5\textwidth]{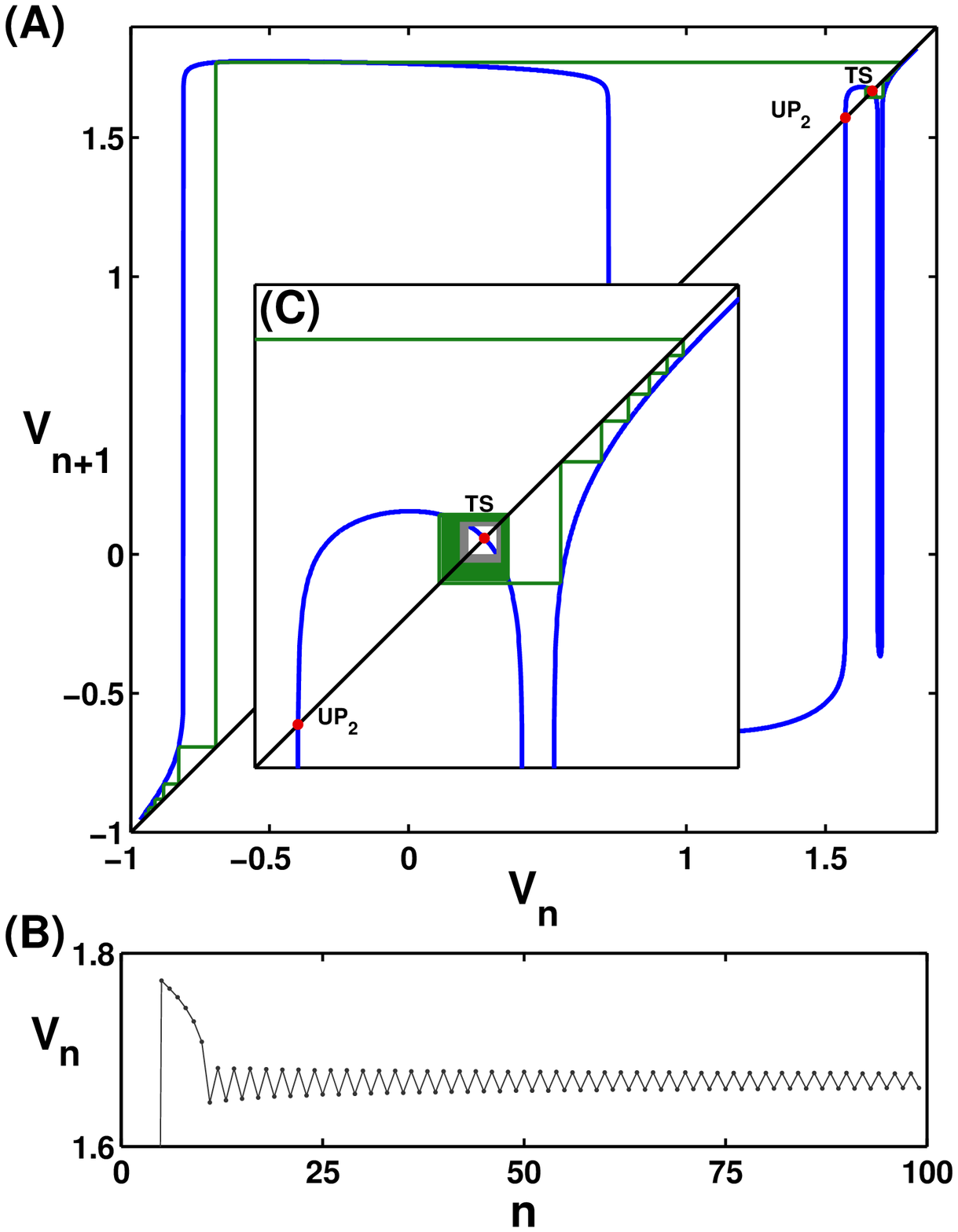}~~\includegraphics[width=0.5\textwidth]{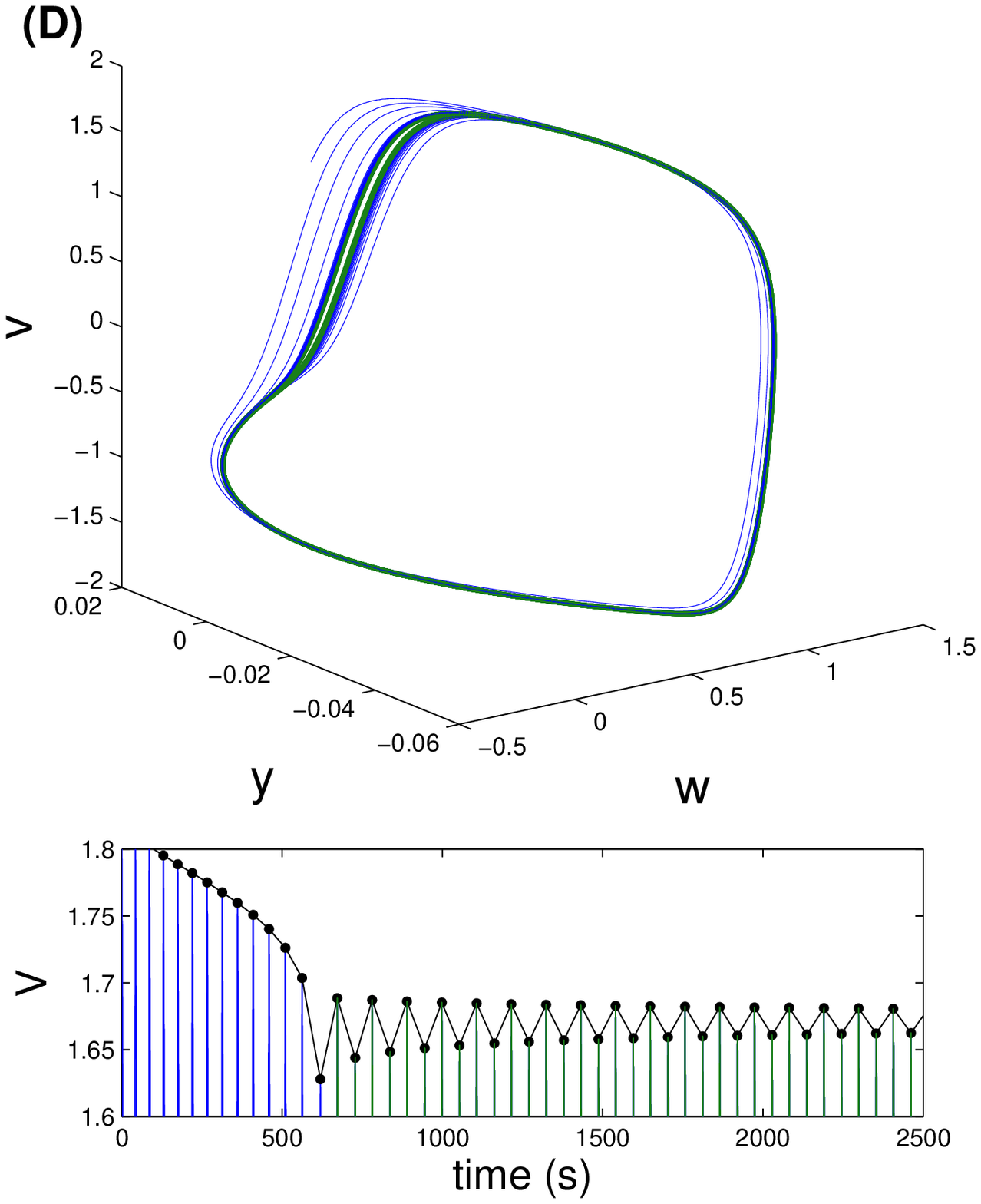}}
\end{center}
\caption{(A) Poincar\'e mapping at $c=-0.6193$ and the voltage trace in (B)
both demonstrate chaotic bursting transients. (C) Enlargement of the right top corner of the mapping shows that
the tonic spiking fixed point has lost the stability through a supercritical period-doubling bifurcation. The new
born period-2 orbit is a new attractor of the mapping, as confirmed  by the zigzagging voltage trace represented in
(B). (D) The same dynamics found directly from integrating the model. We find after a short transient (blue) the model dynamics converge to a period 2 orbit (green)
 as indicated from the mapping (A).}\label{fig6}
\end{figure}

We begin where the model is firmly in the tonic spiking regime at $c=-0.594355$. Tonic spiking is caused by the presence
of a stable periodic orbit located far from the fold on the manifold $\mathrm{M_{lc}}$ (Fig.~\ref{fig1}). The only v-maximum
of this orbit corresponds to a stable fixed point, labeled TS in Fig.~\ref{fig4}(A).
The flat section of the mapping graph adjoining the stable fixed point clearly indicates a rapid convergence to the
point in the v-direction, as shown by the trace in inset (B).   Here the slope of the mapping
reflects the exponential instability (stability) of the quiescent (tonic spiking) branch, made of unstable equilibria and stable limit cycles
of the fast subsystem of the model.

 The formation of the cusp is an indication of a change in
dynamics for the mapping. Thus the mapping insinuates a transition in dynamics of the model~(\ref{fhr})
prior to occupance. Note that the maximal voltage trace provides no indication of any eminent transition in the model's behavior.
The mapping in Fig.~\ref{fig4}(A,\,B), taken for the parameter $c=-0.595$, clearly illustrates that after the cusp has dropped
below the bisectrix, two additional fixed
points, $\mathrm{UP_1}$ and $\mathrm{UP_2}$, are created. $\mathrm{UP_1}$ and $\mathrm{UP_2}$
have emerged through a preceding fold or saddle-node bifurcation taking place at some intermediate parameter
value between $c=-0.594255$ and $c=-0.595$.
Again, let us stress that the singular limit of the model at $\mu=0$  gives a single saddle-node bifurcation through
which the tonic spiking periodic orbit looses stability after it reaches the fold on the tonic spiking manifold.
We point out that, for an instant, the model becomes bistable right after the saddle-node
bifurcation in Fig.~\ref{fig4} leading to the emergence of another stable fixed point with an extremely narrow basin of attraction. Here, as before
the hyperbolic tonic spiking fixed point, TS, dominates the dynamics of the model.

\begin{figure}[b!]
\begin{center}
\centerline{\includegraphics[width=0.4\textwidth]{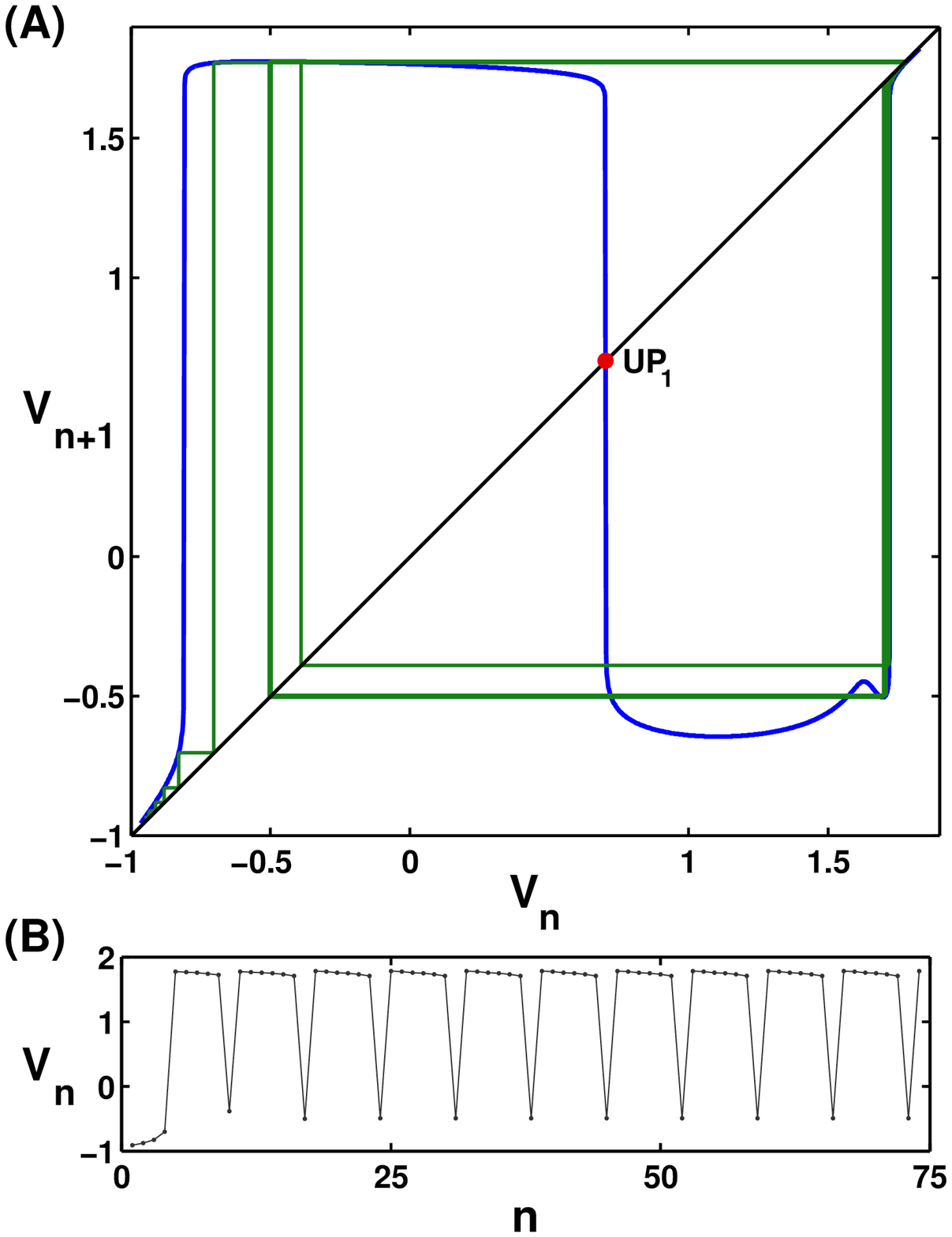}~~\includegraphics[width=0.4\textwidth]{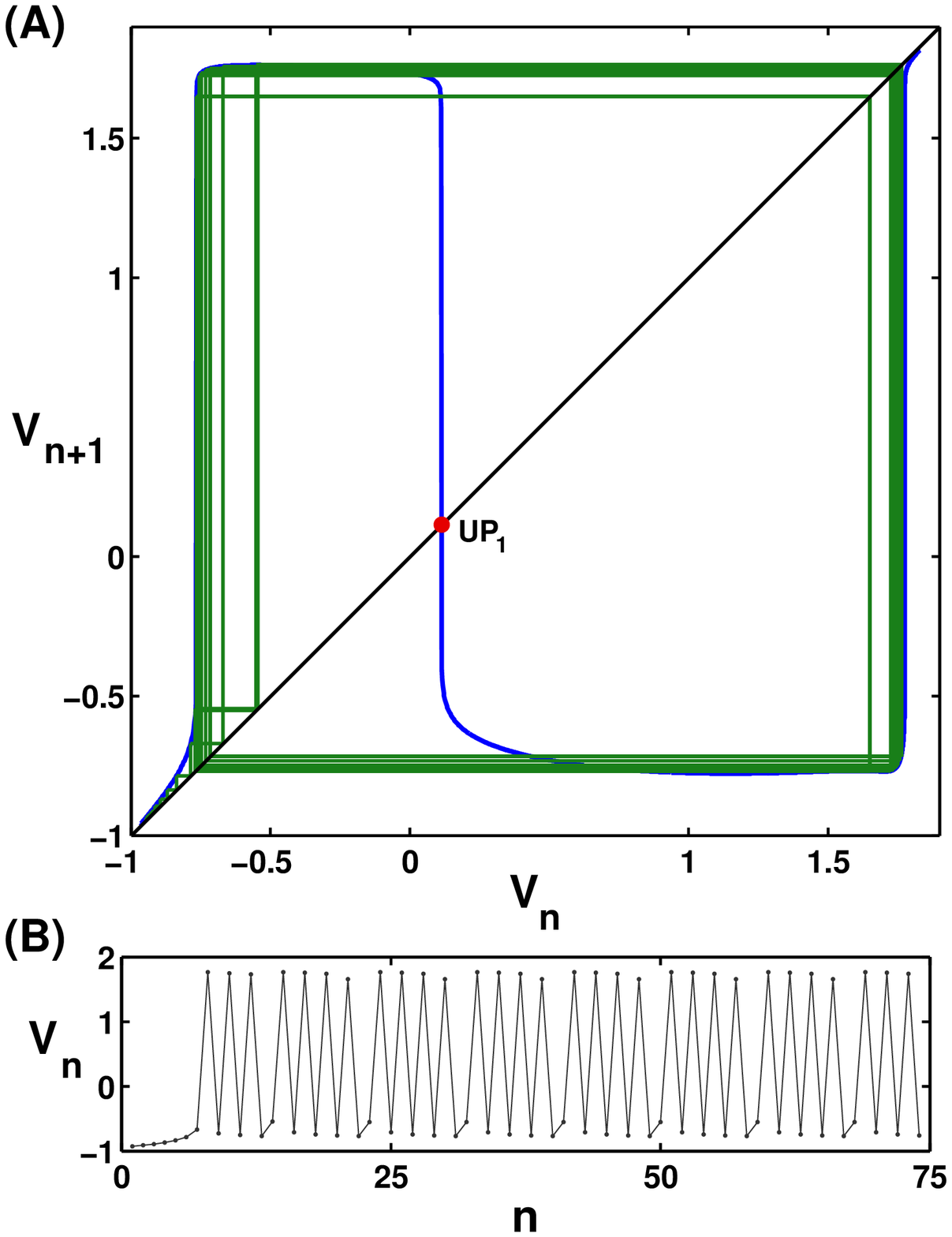}}
\end{center}
\caption{(A) Periodic  bursting with five spikes in the Poincar\'e interval mapping for the FitzHugh-Nagumo-Rinzel model at $c=-0.6215$.
The single unstable fixed point $\mathrm{UP_1}$ separates the tonic spiking section of the mapping from the quiescent or subthreshold section
(left). The number of iterates of the phase point adequately defines the ordinal type of bursting (B). Note a presence of a small
hump around $(\mathrm{V_0}=1.6,\, \mathrm{V_1}=-0.5)$ which is an echo of the saddle-node bifurcation. (C) Poincar\'e return mapping at $c=-0.75$.
Here we find a burst pattern of 3 spikes followed by 2 small amplitude oscillations. The mappings are able to capture all the bursting patterns exhibited by the model.}\label{fig7}
\end{figure}

Figure~\ref{fig5}(A) demonstrates that, as the parameter is decreased further to $c=-0.615$,  the gap between the
new fixed points widens as the point $\mathrm{UP}_2$ moves toward the stable tonic spiking point, TS, indicating a possible saddle-node bifurcation is eminent. Through
this saddle-node bifurcation,  these fixed points merge and annihilate each other; thereby terminate the tonic spiking activity in
the FitzHugh-Nagumo-Rinzel model. Before that happens however, several bifurcations involving
the fixed point, TS,  drastically reshape the dynamics of the model. First, the multiplier becomes negative around
$c=-0.619$, which is the first indication of an impending period doubling cascade.
This is confirmed by the mapping at $c=-0.6193$ in Fig.~\ref{fig6}(A,\,B\,C) showing
that the fixed point has become unstable through the supercritical period-doubling bifurcation. Furthermore, the dynamics of the mapping is
directly mimicked in the full model behavior, see Fig~\ref{fig6}(D)

The new born period-2 orbit becomes the new tonic spiking attractor of the mapping. Observe
from the voltage trace in Fig.~\ref{fig6}(B)
the long transient bursting behavior thus indicating that boundaries of the attraction basin of the period-two orbit become fractal.
Next, the model approaches bursting onset. Correspondingly, the FitzHugh-Nagumo
Rinzel model starts generating chaotic trains of bursts with randomly alternating numbers of spikes per burst.
The number of spikes depends on how close the trajectory of the mapping comes to the unstable
(spiraling out) fixed point, TS, that is used to represent the tonic spiking activity. Each spike train is interrupted  by a single
quiescent period.
The unstable point, $\mathrm{UP_1}$, corresponds to a saddle periodic orbit of the model that is
located on the unstable, cone-shaped section of the tonic spiking manifold $\mathrm{M_{lc}}$ in Fig.~\ref{fig1}.
Recall that this saddle periodic orbit is repelling in the fast variables and stable in the slow
variable.

By comparing Figs.~\ref{fig4}-\ref{fig7} one could not foresee that the secondary saddle-node bifurcation
eliminating the tonic spiking fixed point TS, or corresponding round stable periodic orbit on the manifold $\mathrm{M_{lc}}$
would be preceded by a dramatic concavity change in the mapping shape causing a forward and inverse cascade of  period doubling
bifurcations right before the tonic spiking orbit TS.
The corresponding fixed point, TS, becomes stable again through a
reverse sequence of period doubling bifurcations before annihilating through the secondary saddle-node
bifurcation. However, the basin of attraction becomes so thin that bursting begins to dominate  the
bi-stable dynamics of the model.  Note that the bursting behavior becomes regular as the phase
points pass through the upper section of the mapping tangent to the bisectrix. The number of
iterates that the orbit makes here determine the duration of the tonic spiking phase of bursting and is followed by a
quiescence period initially comprised of a single iterate of the phase point to the right of the threshold $\mathrm{UP_1}$.
The evolution of bursting into MMO and on to subthreshold  oscillations will be discussed in the next section.

\subsection{From bursting to mixed-mode oscillations and quiescence}\label{mmo}

\begin{figure}[b!]
\begin{center}
\centerline{\includegraphics[width=0.5\textwidth]{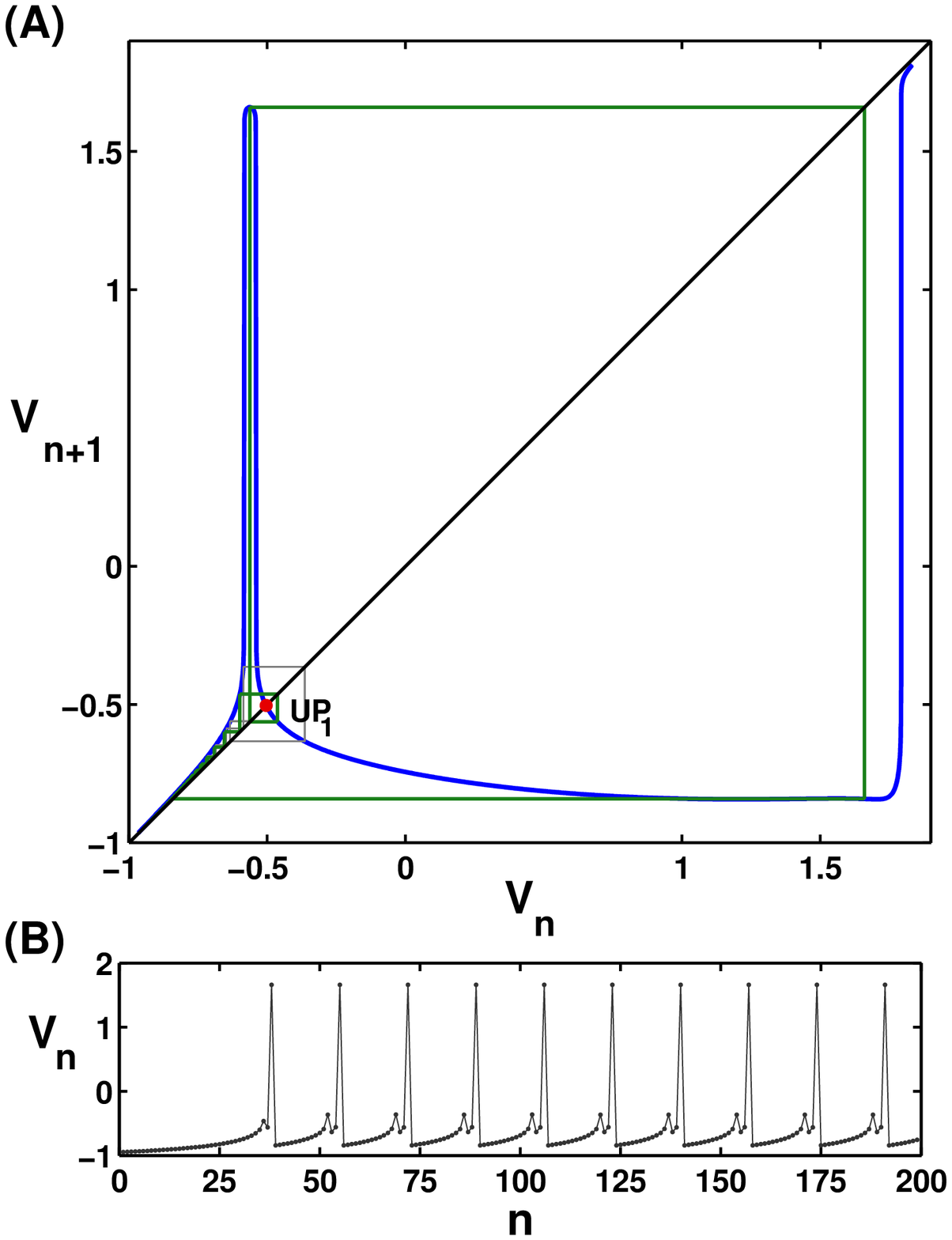}~~\includegraphics[width=0.48\textwidth]{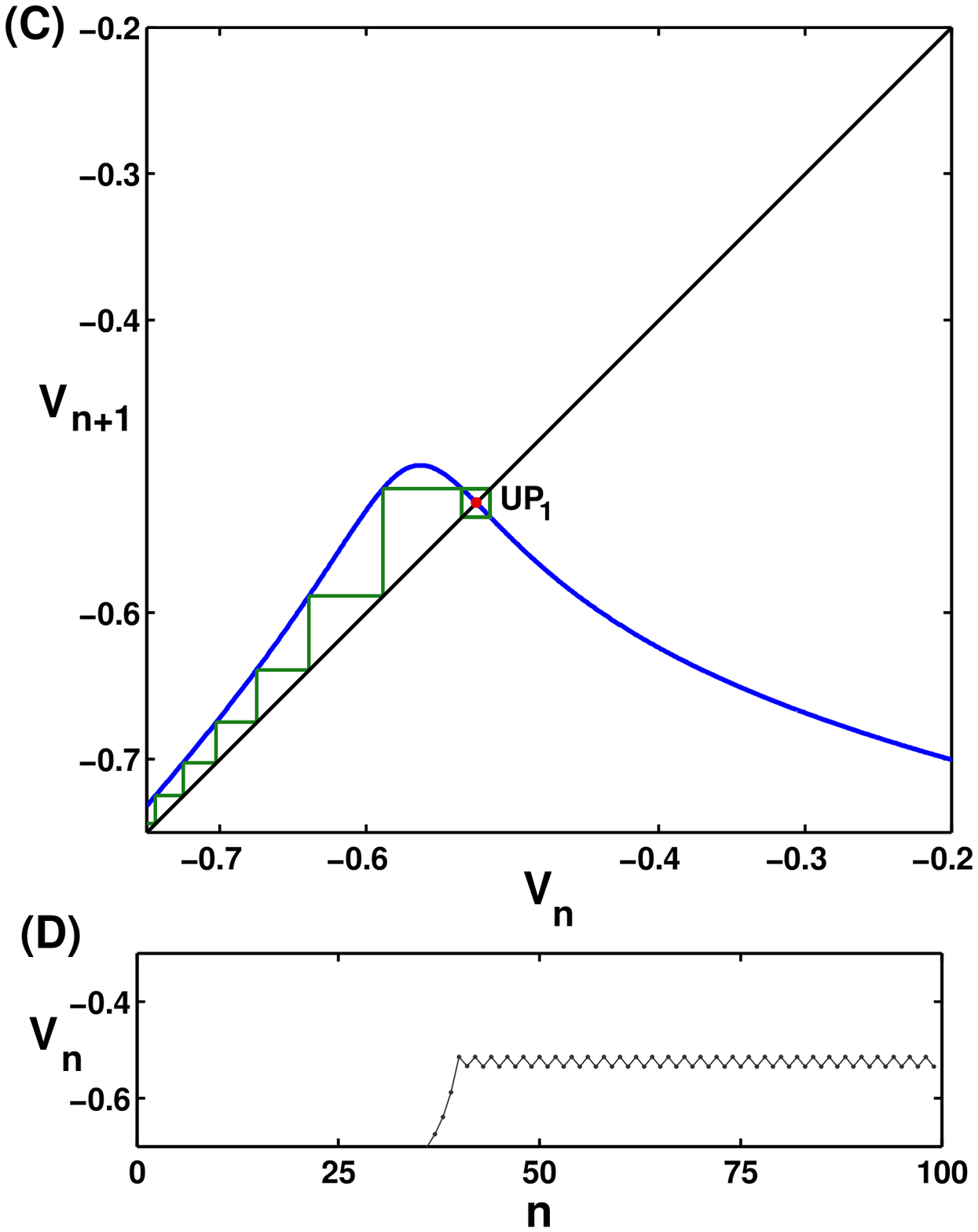}}
\end{center}
\caption{(A) Chaotic MMO and bursting in the mapping  at $c=-0.904$ caused by the complex recurrent behavior around the unstable
fixed point $\mathrm{UP_1}$. (B) Subthreshold oscillations are disrupted sporadically by large and intermediate magnitude spikes
thereby destroying the rhythmic bursting  in the model. (C) Poincar\'e return mapping for the FitzHugh-Nagumo-Rinzel model shows
no bursting but complex subthreshold period 2 oscillations at $c=-0.908$. (D) After the peak in the mapping decreases in amplitude, high amplitude spikes becomes impossible.  Here, chaos is caused by homoclinic
orbits to the unstable fixed point $\mathrm{UP_1}$, just prior to this figure. }\label{fig8}
\end{figure}

The disappearance of the tonic spiking orbit, TS, accords with the onset of regular bursting in the mapping and in the
 model (\ref{fhr}). In the mapping, a bursting orbit is comprised of iterates on the tonic spiking and quiescent sections
separated by the unstable threshold fixed point, $\mathrm{UP_1}$, of the mapping in Fig.~\ref{fig7}.
 The shape of the graph undergoes a significant change reflecting  the change
in dynamics.  The fixed points in the upper right section of the mapping disappear through
a saddle-node bifurcation. One of the features of the saddle-node is the bifurcation memory: the phase point continues to
linger near a phantom of the disappeared saddle-node. The mapping near the bisectrix can generate a large number of iterates before
the phase points diverge toward the quiescent phase. The larger the number of iterates near the
bisectrix corresponds to a longer tonic spiking phase
of bursting.  Figure \ref{fig7} demonstrates how the durations of the phases change along with a change in the mapping shape: from a
single quiescent iterate to the left of the threshold, $\mathrm{UP_1}$, to a single tonic-spiking iterate corresponding to a bursting
orbit with a single large spike  in  the model.

The transition from bursting to quiescence in the model is not monotone because the regular
dynamics may be sparked by episodes of chaos.
Such subthreshold chaos in the corresponding mapping at $c=-0.9041$ is demonstrated in Fig.~\ref{fig8}(A).
This phenomena is labeled MMO because the small amplitude subthreshold oscillations are
sporadically interrupted by larger spikes (Inset B).
Use of the mapping makes the explanation of the phenomena in elliptic bursters particularly clear.
In Fig.~\ref{fig8}(A), after the mapping (or the model) fires a spike, the phase point is reinjected close to the threshold point,
$\mathrm{UP_1}$, from where it spirals away to make another cycle of bursting. Note that the number of iterates of the phase point
around $\mathrm{UP_1}$ may vary after each spiking episode.   This gives rise to solutions that are called bi-asymptotic or
homoclinic orbits to the unstable fixed point $\mathrm{UP_1}$. The occupancy of such a homoclinic orbit to a repelling
fixed point is the generic property of a one-dimensional non-invertible mapping \cite{Mira1987}, since the point of a
homoclinic orbit might have two pre-images.  Note that the number of forward iterates of a homoclinic point may be  finite in a
non-invertible mapping, because the phase point might not converge, but merely jump onto the unstable fixed point after being reinjected.
However, the number of backward iterates of the homoclinic point is infinite, because the repelling fixed point
becomes an attractor for an inverse mapping in restriction to the local section of the unimodal mapping, see Fig.~\ref{fig8}(A,\,B).
The presence of a single homoclinic orbit leads to the abundance of other emergent homoclinics \cite{Gavrilov1972} via a homoclinic explosion
\cite{Shilnikov2001}.

\begin{figure}[b!]
\begin{center}
\centerline{\includegraphics[width=0.35\textwidth]{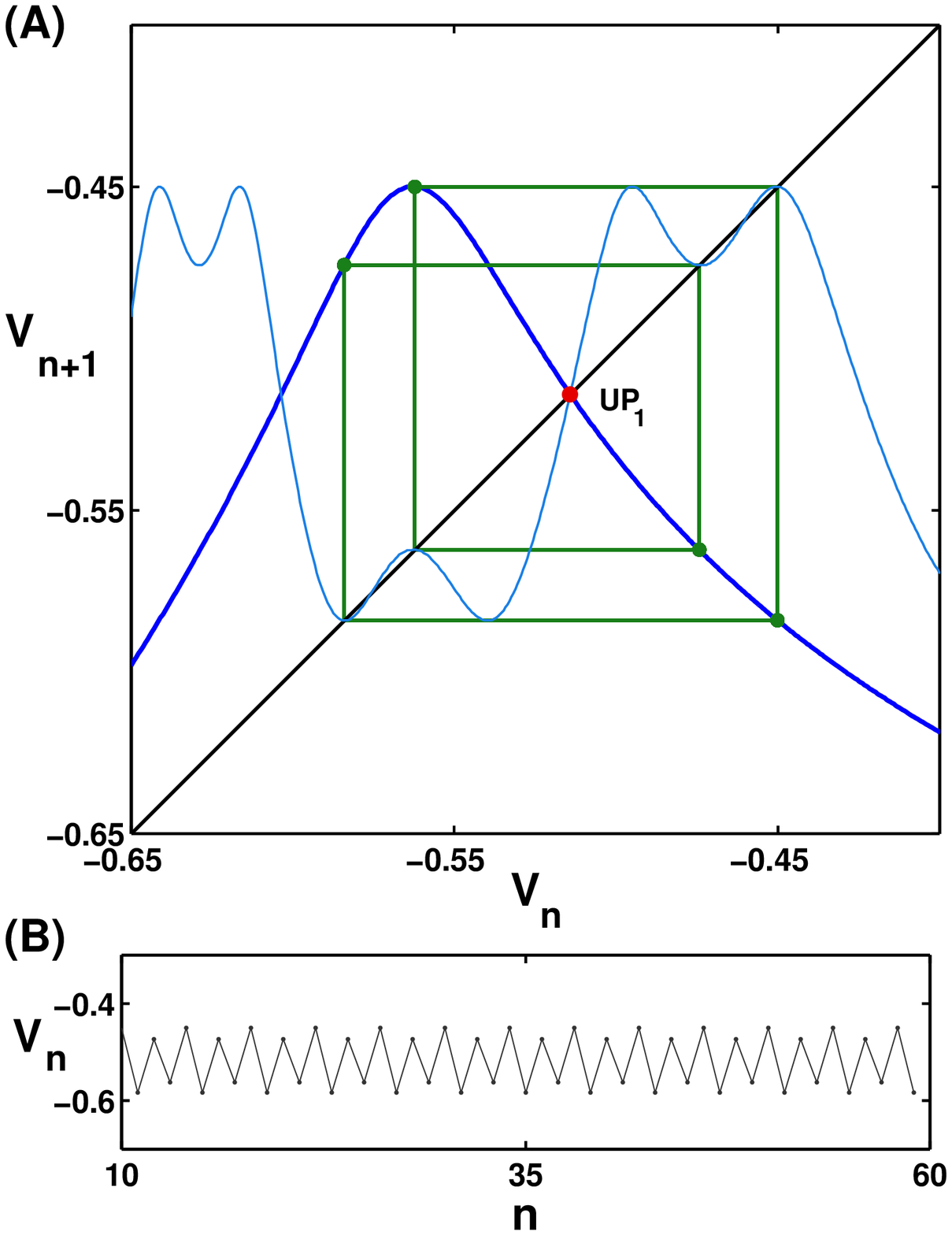}~~\includegraphics[width=0.35\textwidth]{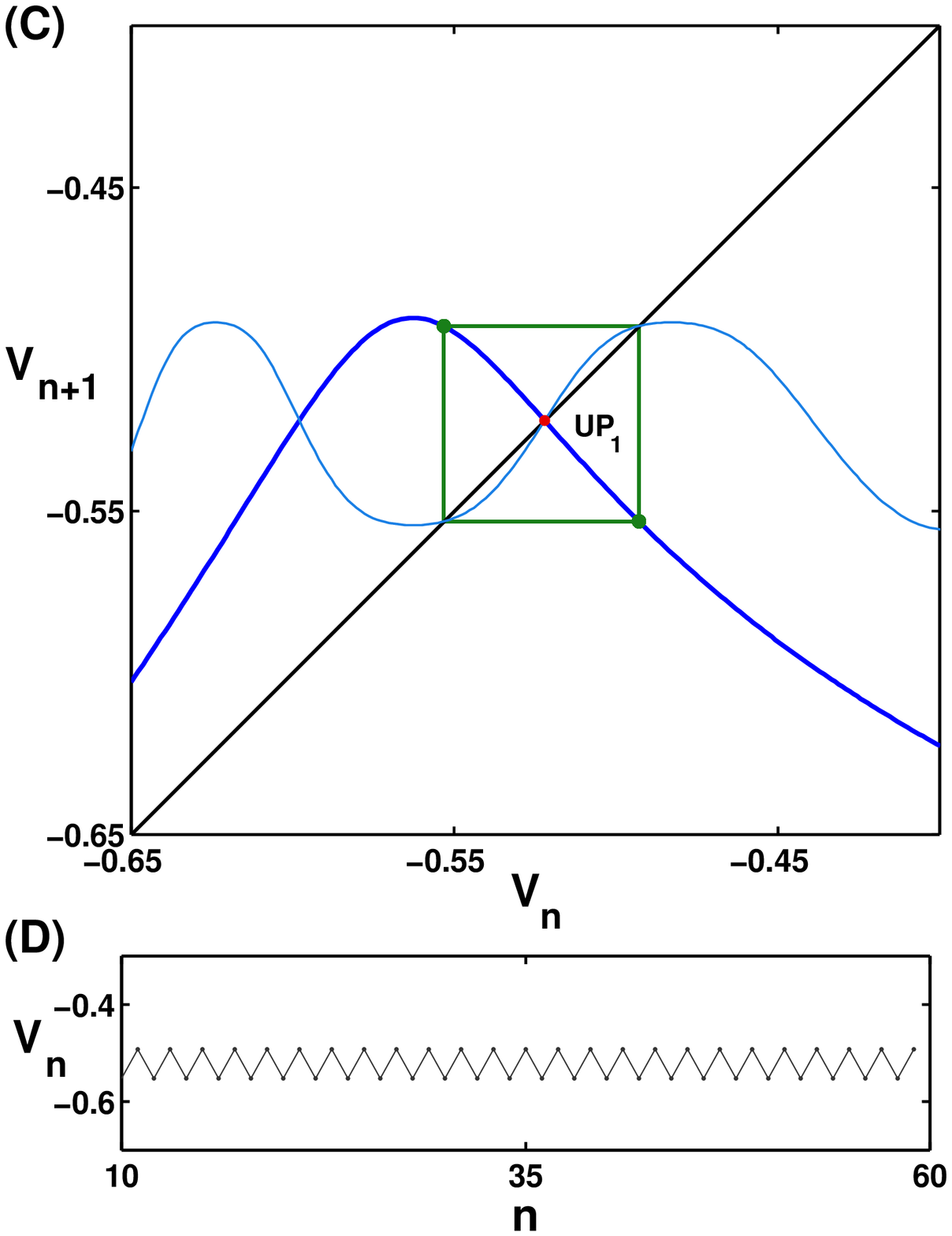}}
\end{center}
\caption{(A) and (C) Show stable period-4 and period-2 orbits (green) of the interval mapping at $c=-0.906$ and
$c=-0.9075$. Shown in light-blue are the corresponding mappings $T^4$ and $T^2$ of degree four and two with four and two
stable fixed points correspondingly.  The traces of the orbits are shown in Insets (B) and (D).}\label{fig9}
\end{figure}

A small decrease of the bifurcation parameter causes a rapid change in the shape of the mapping, as depicted in Figs.~\ref{fig8}(C,\,D). The sharp peak near the threshold becomes lower so that the mapping can no longer generate large amplitude spikes.
As the parameter is decreased further, the unstable fixed point, $\mathrm{UP_1}$, becomes stable through
a reverse period-doubling cascade.
The last two stages of the cascade are depicted in Fig.~\ref{fig9}. Insets~(A) and (C) of the figure show
 stable period-4 and period-2 orbits, and their traces in Insets~(B) and (D),
 as the parameter $c$ is decreased from -0.906 to $-0.9075$. Here we  demonstrate another ability of the interval mappings derived directly from the flow. In addition
to the original mapping, T, in Fig.~\ref{fig9} we see two superimposed mappings, $T^2$ and $T^4$, (shown in light blue) of degree two and four respectively. The four points of periodic orbit in Inset~(A) corresponds to the four fixed points of the
fourth degree mapping $T^4$ at $c=-0.9075$,
whereas the period-two orbit in (C) correspond to two new fixed points of the mapping $T^2$ in (C) at $c=-0.9075$. We see clearly that
both periodic orbits are indeed stable because of the slopes of the mappings at the fixed points on the bisectrix. Using the mappings of higher
degrees we can evaluate the critical moments at which the period-two and period-four orbits are about to bifurcate.
We point out that a period-doubling cascade, beginning with a limit cycle near the  Hopf-initiated canard toward subthreshold chaos has been
 recently reported in slow-fast systems \cite{Zaks2005,Zaks2011}.

Decreasing $c$ further, the period-two orbit collapses into the fixed point, $\mathrm{UP_1}$, which
becomes stable. The multiplier, first negative becomes positive  but is still less than one in the absolute value.
 In terms of the model, this means that the periodic orbit collapses into a
saddle-focus through the subcritical Andronov-Hopf
bifurcation. After that, the equilibrium state, located at the intersection of the manifold $\mathrm{M_{eq}}$ with the slow-nullcline (plane)
in Fig.~\ref{fig1}, becomes stable and the model goes into quiescence for  parameter values smaller then  $c=-0.97$.
The stable equilibrium state corresponds to the fixed point, Q,  which  is the global attractor in the mapping.

\section{Quantitative features of mappings: Kneadings}\label{quan_map}

\begin{figure}[b!]
\begin{center}
\centerline{\includegraphics[width=0.75\textwidth]{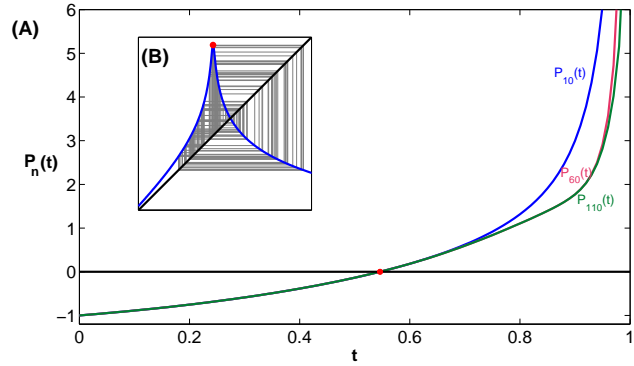}}
\end{center}
\caption{(A) Graphs of the three polynomials, $P_{10}(t)$,  $P_{60}(t)$ and $P_{110}(t)$ defined on the unit interval,
and generated through the series of the signed kneadings at  $c=-0.90476$.  Inset (B) shows the corresponding interval mapping.
The iterates of the critical point, $v_c$, determine the symbolic dynamics for the unsigned kneading symbols: $-1$ if the the phase
point lands on the decreasing section of the mapping graph to the right of the critical point, and $+1$ if it lands to the increasing
section of the mapping, which is to the left of the critical point.}\label{fig10}
\end{figure}

In this section we discuss a quantitative property of the interval mappings for
the dynamics of the model (\ref{fhr}). In particular, we carry out the examination of complex dynamics
with the use of calculus-based and calculus-free tools such as Lyapunov exponents and kneading invariants for the
symbolic description of MMOs.

Chaos may be  quantitatively measured by a Lyapunov exponent. The Lyapunov exponent is evaluated for the one-dimensional mappings as follows:
\begin{equation}
\lambda=\lim_{N \to + \infty} \frac{1}{N} \sum_{i=1}^{N} \log| T^\prime(v_{i})|,
\end{equation}\label{lyap}
where $T'(v_i)$ is the slope (derivative) of the mapping at the current iterate $v_i$ corresponding to the $i$-th
step for $i=0,\dots,N$. Note that, by construction, the mapping graph is a polygonal and to accurately evaluate the derivatives
in (\ref{lyap}) we used a cubic spline.
The Lyapunov exponent, $\lambda$, yields a lower bound for the topological entropy $h(T)$ \cite{katok}; serving as a measure of chaos
in a model. The Lyapunov exponent values $\lambda \simeq 0.24$ and $\lambda \simeq 0.58$, found for the interval mappings at $c=-0.9041$ and
$c=- 0.90476$, resp., show that chaos is developed more in the case of subthreshold oscillations than for MMOs.

The topological entropy may also be evaluated though a symbolic description of the dynamics of the mapping that
require no calculus-based tools. The curious reader is referred to \cite{Glen1996,Malkin2003}   for the in-depth
and practical overviews of the kneading invariants, while below we
will merely touch the relevant aspects of the theory. For unimodal mappings of an interval into itself with a single critical point, $v_c$,
like for the case $c=-0.90476$, we need only to follow the forward iterates of
the critical point to generate the {\em unsigned kneading sequence}
$\kappa(v_c)=\{ \kappa_n(v_c) \} $ defined on $\{-1,\,+1\}$ by the following rule:
\begin{equation}
\kappa_n (v_q) =  \left\{
\begin{array}{cc}
  +1, &  \mbox{if~~} T^n(v_c) < v_c \\
-1, &  \mbox{if~~} T^n(v_c) > v_c;
\end{array}\right.
\end{equation}
here $T^n(v_c)$ is the n-th iterate of the critical point $v_c$.

The kneading invariant of the unimodal mapping is a series of the {\em signed kneadings} $\{ \tilde{\kappa}_n \}$ of the critical
point, which are defined through the unsigned kneadings, $\kappa_i$, as follows:
\begin{equation}
\tilde{\kappa}_n =  \prod_{i=1}^{n} \kappa_i,
\end{equation}
or, recursively:
\begin{equation}
\tilde{\kappa}_n = \kappa_n \, \tilde{\kappa}_{n-1}, \qquad i=2,3,...\,.
\end{equation}
Next we construct a formal power series;
\begin{equation}
P(t)=\sum_{i=0}^\infty \tilde{\kappa}_{i}\,t^i.
\end{equation}
The smallest zero, $t^*$ (if any), of the series within an interval $t\in (0,\,1)$ defines the topological entropy, $h(T) = \ln(1/t^*)$.
The sequence of the signed kneadings, truncated to the first ten terms,
$
\{ - \,  + \,  +  \,  +  \, - \,  + \, + \, + \, - \, + \}
$
for the mapping in Fig.~\ref{fig10} inset~B, generates
the polynomial $P_{10}(t)=-1 +t + t^2 + t^3 - t^4 + t^5 + t^6 + t^7 - t^8 + t^9$. The single zero of $P_{10}(t)$ at
$t^* \approx 0.544779$ yields a close
estimate for the topological entropy $h(T) \approx 0.6073745$, see Fig.~\ref{fig10}(A).
The advantage of an approach based on the kneading invariant to quantify chaos
is that evaluation of the topological entropy does not involve numerical calculus for such equationless
interval mappings, but relies on the mixing properties of the dynamics instead. Moreover, it requires relatively
few forward iterates of the critical point to compute the entropy accurately,
as the polynomial graphs in Fig.~\ref{fig10} suggests. Besides yielding the quantitative information such as the topological
entropy, the symbolic description based on the kneading
invariants provide qualitative information for identifying the corresponding Farey sequences describing the MMOs in terms
of the numbers of subthreshold and tonic spiking oscillations.

\section{Discussion}

We present a case study for an in-depth examination of the bifurcations that take place at activity transitions between
tonic spiking, bursting and Mixed Mode Oscillations in the FitzHugh-Nagumo-Rinzel model. The analysis is accomplished
through the reduction to a single-parameter family of equationless Poincar\'e  return mappings for an interval of the
 ``voltage" variable. We stress that these mappings are \emph{models} themselves for evaluating the complex dynamics
 of the full three-dimensional model. Nevertheless, the dynamics of the single
accumulative variable, $v$, reflects the cooperative dynamics of other variables in the model.
The reduction is feasible since the model is a slow-fast system and, hence, possesses a two-dimensional,
slow-motion tonic-spiking manifold around which the oscillatory solutions of the models linger.
We have specifically concentrated on the dynamics of the voltage \cite{Channell2007a,Channell2009}, as it is typically the only
measurable, and thus comparable,  variable in experimental studies in neuroscience and physical chemistry.

It is evident that no 1D return mapping of the interval is intended to detect a torus in the phase plane,
whereas the pointwise mappings generated by a forward
time series of the voltage can identify the torus formation in the phase space. Note that the torus has a canard-like nature, that is
the torus exists within a narrow parameter window. A torus formation in a 3D model with two slow variables near the fold  was reported
also in \cite{torus2slow}. Another parallel of the FitzHugh-Nagumo-Rinzel model with electrochemical systems, including the Belousov-Zhabotinky
reaction, is that the latter also demonstrates a quasiperiodic regime \cite{argoul}. The emergence of the torus near the fold of the
tonic spiking manifold first described in \cite{Shilnikov2003,Cymbalyuk2005a} has turned out to be a generic phenomenon observed recently in
several plausible models \cite{kuzn,lp2011}, including a model for the Purkinje cells \cite{Kramer2008a,burke}, and in a 12D hair cell model \cite{Neiman2011}

A minor drawback of the approach is a small detuning offset in
parameter values at which the model and the mapping have nearly the same
dynamics, matching orbits, or undergo the same bifurcations. This is caused by the fact that a one-dimensional mapping
for a single voltage variable does not fully encompass the dynamics of other, major and minor, variables of the corresponding model.
In general, most features of a dissipative model with a negative
divergence of the vector field that results in a strong contraction of the phase volumes, are adequately modeled by a 1D Poincar\'e
mapping. However, this is not true when such a contraction is no longer in place, for example, when the divergence becomes sign-alternating.
There are two such places near the manifold $\mathrm{M_{lc}}$ in the model~(\ref{fhr}): one  is near the fold, the
second is close to the cone-shaped tip.  The sign alternating near the tip of the cone is  where the model has an equilibrium state of the saddle-focus
type with a pair of complex conjugate eigenvalues with a small positive real part and a real negative eigenvalue due to the
Andronov-Hopf bifurcation and the smallness of  $\varepsilon$.

The algorithm for interval mapping construction has two stages. First, one needs to identify the tonic spiking manifold
in the phase space of the slow-fast neuron model in question. This is accomplished by either using the geometric dissection method,
or the parameter continuation technique.
The more accurately and completely the first stage is performed the more natural and smooth these numerically derived mappings will be.
The second stage is to build the mappings for a range of parameter values.  The analysis of such mappings lets one
identify not only attractors, but more importantly, the unstable sets including fixed, periodic and homoclinic orbits,
which are known to be the globally organizing centers governing the dynamics of any model.
In addition, having computationally smooth mappings allows one to create symbolic descriptions for dynamics, compute kneading
invariants, evaluate Schwarzian derivatives etc., as well as estimate other quantities measuring the degree of complexity for
the trajectory behavior like  Lyapunov exponents and topological entropy.

Our computational method allows us to thoroughly describe the bifurcations that the model (\ref{fhr}) undergoes while
transitioning between  states: from tonic spiking to bursting and then to quiescence. Taken individually,
each mapping offers only a glimpse into the system behavior. However, with an entire family of mappings we obtain
deep insight into the evolution of the model's dynamics though the interplay and bifurcations of the fixed points
and periodic orbits. This allows for not only the description of  bifurcations {\em post factum}, but
to predict the changes in the dynamics of the model under consideration  before they actually occur. For additional analysis on elliptic
bursters including torus formation, we refer the reader to \cite{wojcik_1d}.


\end{document}